# Competing *s-p* and *p-p* fluctuations in charge-disproportionation of BaBiO$_3$

Sumit Sarkar, Ram Janay Choudhary* and Rajamani Raghunathan*

*UGC-DAE Consortium for Scientific Research, DAVV Campus, Khandwa Road, Indore 452001, India*
*Corresponding authors: ram@csr.res.in and rajamani@csr.res.in*

**Abstract:** Here we investigate the mechanism of charge-disproportionation (CD) in BaBiO$_3$ (BBO) using density functional theory under different crystal symmetries and by employing strain as an external perturbation. The competition between Bi 6sp-O 2p (s-p) and O 2p-O 2p (p-p) charge-fluctuations decides the electronic ground state, charge-disproportionation and bond-disproportionation (BD) in BBO. An extended Hubbard Hamiltonian involving onsite (U) and long-range (V) coulomb repulsions is also employed to ascertain the microscopic conditions for the formation of the lone-pair on bismuth site. A strong tensile strain increases p-p fluctuation and enhances negative-CT character, while strong compressive strain favors s-p fluctuation leading to more positive-CT character. Bulk BBO is at the boundary of positive and negative-CT regimes.

Charge ordering (CO) in transition metal oxides (TMOs) has been an interesting topic of research for the past few decades [1-4]. However, the presence of magnetic interactions hinders the understanding of CO in these oxides. ABO$_3$ perovskites containing A=Sr, Ba or Ca and B=Bi have recently been engrossing scientists due to their emergent properties like the superconductivity under hole doping, topological insulator, two-dimensional electron gas and charge density wave state [5-12]. In these compounds, the Bi ion does not stabilize in its formal 4+ valence state, but forms two distinct charge states at alternate Bi sites, known as the charge-disproportionation (CD) [10,13-15]. CD has been studied theoretically in the context of competing on-site ($U$) and inter-site ($V$) $e$-$e$ correlations, as well as by introducing a negative-U parameter [16,17]. Few studies have also suggested that the system can possess bond disproportionation (BD) without CD on the Bi site, in contrast to the previous neutron and x-ray diffraction and spectroscopic studies [11,13,18-21].

Considering electron-electron (*e-e*) and electron-phonon (*e-ph*) interactions, the mechanism of CD in BaBiO$_3$ (BBO) is generally debated between a purely ionic (Bi$^{3+}$ - Bi$^{5+}$) and covalent (Bi$^{4-\delta}$ – Bi$^{4+\delta}$ (or) $Bi^{3+}\underline{L} \rightarrow Bi^{3+} + Bi^{3+}\underline{L}^2$) limits, with $\underline{L}$ representing a ligand hole, [11,13,20-23]. Using a molecular orbital (MO) picture, it has recently been shown that the CD is a result of dynamic fluctuation between Bi 6sp – O2p *and* Bi 6p – O 2p hybridizations driven by the octahedral breathing phonon mode [14]. A previous pressure dependent density functional theory (DFT) study has shown that the CD can persist up to 100 GPa and no insulator-metal transition (IMT) emerges [24-26]. Despite such studies, why BBO does not show IMT and whether BBO is a positive or negative charge transfer (CT) insulator, particularly in the context of hybridization fluctuation still remains unexplored. As the CD of BBO can be tuned by controlling the Bi – O hybridization, its relation with lattice strain and symmetry is also unknown. Another open question is, are the CD and BD correlated? The insights from these studies can enable a new avenue to tune the CD by controlling the hybridization and also provide a good starting point to study CO in more complex oxides like ferrites, cobaltates and nickelates [1-4].

In this Letter, we focus on the role of strain and crystallographic symmetry on the electronic structure and CD mechanism of BBO by controlling the Bi $6sp$ – O $2p$ and Bi $6p$ – O $2p$ type hybridization using DFT [27-35] calculations and x-ray photoemission spectroscopy. We show that the dominance of one of these hybridization over the other governs the charge state of a bismuth site, further leading to uncharacteristic CD. Our study also establishes the conditions at which BBO shows positive- or negative-CT insulating character.

*Crystal structure.-* BBO exists in four structural phases, with space groups *(a) I2/m, (b) P2$_1$/n, (c) R$\bar{3}$ and (d) Fm$\bar{3}$m* as shown in fig. S1 and S2. The structural evolutions of BBO are related to the modification of octahedral tilting distortion (OTD) and octahedral breathing distortion (OBD) present in the system. The *I2/m* (*P2$_1$/n*) structure shows two (three) different Bi - O - Bi bond angles, whereas the $R\bar{3}$ and $Fm\bar{3}m$ structures show only one unique bond angle. This is because of the different tilting patterns present in the corresponding space group [36-39]. All the non-cubic structures show large deviation of Bi - O - Bi bond angles from 180° indicative of a strong OTD. In *P2$_1$/n* case the Bi - O - Bi angle along the *c*-axis ($\phi_3$) is different from that in the *ab*-plane ($\phi_1$, $\phi_2$), but for the *I2/m* case $\phi_3=\phi_1$ (fig. S1).

The variation of bond angle ($\phi$) with applied strain for monoclinic (*I2/m* and *P2$_1$/n*) and rhombohedral ($R\bar{3}$) structures is shown in fig.1 (a). All the bond angles increase linearly as we go from -2% to +2% strain in all non-cubic space-groups. Though both the monoclinic structures belong to the same point group, the spread of $\phi$ is much larger in *I2/m* than in *P2$_1$/n*. The cubic phase does not possess any OTD and the bond angle remains at 180° [40]. These findings imply that the covalency of Bi - O could be different in different planes resulting in a non-uniform distribution of the hole density in the O - $2p$ band as shown in fig. S3 for the *P21/n* phase.

We also note that in BBO, the alternate Bi - O bonds disproportionate into long and short bonds corresponding to bismuth ion with lower and higher charge states respectively, known as BD or dimerization. BD is quantified by the difference in these two bond lengths, $\delta L$, whose strain dependence is shown in Fig.1 (b). The BD as well as the actual bond lengths of both the monoclinic and rhombohedral structures are more or less the same. In all the four structures, the BD increases from -2% to +2% strain.

*Many-body picture of CD.-* The MO picture presented in an earlier work (fig. S4) showed that, in case of Bi $6p$ – O $2p$ hybridization, the Bi $6s$ orbital does not hybridize with the O $2p$ states and forms a lone pair of electrons. When both Bi $6s$ and $6p$ take part in hybridization with O $2p$ to stabilize 5+ charge state, the energy of the $\sigma$ MO is considerably lowered. It is puzzling why a lone-pair of electrons prefer to occupy a high energy 6s orbital, when a hybridized orbital of lower energy is available. In order to understand this, we construct a rudimentary model involving only the 6s orbital of two bismuth atoms with one orbital and one electron per bismuth site ($6s^1$) to describe charge disproportionation.

$$\hat{H} = -\sum_{\sigma=\uparrow,\downarrow} t(\hat{a}^\dagger_{1,\sigma} a_{2,\sigma} + \hat{a}^\dagger_{2,\sigma} a_{1,\sigma}) + \varepsilon_1 \hat{n}_1 + \varepsilon_2 \hat{n}_2 + \frac{1}{2} U_1 \hat{n}_1(\hat{n}_1 - 1) + \frac{1}{2} U_2 \hat{n}_2(\hat{n}_2 - 1) + V_{12} \hat{n}_1 \hat{n}_2$$

$$= -\sum_{\sigma=\uparrow,\downarrow} t(\hat{a}^\dagger_{1,\sigma} a_{2,\sigma} + \hat{a}^\dagger_{2,\sigma} a_{1,\sigma}) + \varepsilon \hat{n}_1 + (\varepsilon - \Delta)\hat{n}_2 + \frac{1}{2} U_1 \hat{n}_1(\hat{n}_1 - 1) + \frac{1}{2} U_2 \hat{n}_2(\hat{n}_2 - 1) + V_{12} \hat{n}_1 \hat{n}_2 \quad \ldots (1)$$

In the above Hamiltonian (1), the first term corresponds to the kinetic energy, with transfer integral parameter *t*. As this term involves hopping of an electron between sites *1* and *2*, it is off-diagonal in nature. The operator $\hat{a}^\dagger_{1,\sigma}(\hat{a}_{1,\sigma})$ creates (annihilates) an electron with spin $\sigma$ on site *1* (and similarly for site *2*). The second and third terms correspond to the site energy of orbitals *1* and *2*. As the hybridization of Bi 6s/6p orbital with O 2p stabilizes the energy of $\sigma$ MO, the *1* and *2* sites can be described by different site energies, $\varepsilon_1$ and $\varepsilon_2$ respectively. The site energy of site *1* can be fixed to a reference value of $\varepsilon$ and that of site *2* can be written as $\varepsilon_2 = \varepsilon - \Delta$, where $\Delta$ is the site energy difference between sites *1* and *2* with the condition $\Delta > 0$. The number operator $\hat{n}_i = \sum_\sigma \hat{a}^\dagger_{i,\sigma} \hat{a}_{i,\sigma}$ where *i=1* or *2* is the site index. The last two terms correspond to the on-site and inter-site (long-range) Coulomb repulsions.

It would be instructive to understand the *t*=0 behaviour of the Hamiltonian, in which case the electronic Hamiltonian is diagonal in nature. For the sake of simplicity, we fix $\varepsilon = 0$ as the reference and the orbital energy of the site *2* is taken to be -$\Delta$. In the singlet space, the basis set consists of three configurations viz: (1) the second site is doubly occupied, (2) both the sites are singly occupied with electrons of opposite spins and (3) the first site is doubly occupied (Table 1). The energies of these three configurations can be readily written as $E_1 = U_2 - 2\Delta$, $E_2 = V - \Delta$ and $E_3 = U_1$. It can be seen that the ground state electronic configuration is decided by the relative strengths of electron correlation parameters, $\Delta$, $U_1$, $U_2$ and *V*. Double occupancy of the lower orbital will be the ground state if $U_2 - V < \Delta$ and $U_2 - U_1 < 2\Delta$. The second configuration with one electron per orbital (|2⟩) is favoured if $U_2 - V > \Delta$ and $V - U_1 < \Delta$. According to band theory this half-filled band should be metallic for all non-zero values of *t*. But, BBO is known to be a semiconductor with a band gap of about 0.6 to 0.8 eV [4,7,9,10,31]. Lone pair of electrons on the higher Bi orbital corresponding to state |3⟩ will be favoured if $V - U_1 > \Delta$ and $U_2 - U_1 > 2\Delta$. This state is stabilized only when $U_2$ is significantly larger than $U_1$, since $\Delta$ is positive. This is meaningful because the $Bi^{5+}$ ion is more compact than the $Bi^{3+}$ ion due to the large positive nuclear charge on the former. A previous literature has also shown that the on-site Coulomb repulsion is much higher for $Bi^{5+}$ than for the $Bi^{3+}$ charge state [41,42].

**Table1:** *Electron configuration of different states and conditions for the state to be the ground state in case of two orbital two electron system for the Hamiltonian presented in eq. (3).*

| State | Configuration | Condition |
|---|---|---|
| $\|1\rangle$ | 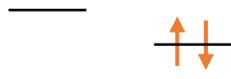 | $U_2 - V < \Delta$ and $U_2 - U_1 < 2\Delta$ |
| $\|2\rangle$ | 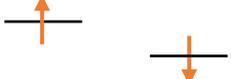 | $U_2 - V > \Delta$ and $V - U_1 < \Delta$ |
| $\|3\rangle$ | 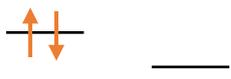 | $V - U_1 > \Delta$ and $U_2 - U_1 > 2\Delta$ |

*Charge-disproportionation.-* In BBO, however, the transfer integral $t \neq 0$ due to the metal ligand hybridization or covalency. In order to test our model and to quantitatively understand the variation of electron number at the two alternate Bi sites (Bi1 and Bi2), we have calculated the Bader charges [43-46] as a function of strain using DFT for different space groups (Table 2). The charges of Bi1 and Bi2 corresponding to two different Wycoff positions are understood by comparing with the corresponding reference states namely, $Bi^{3+}$ (13.10e), $Bi^{4+}$ (12.77e) and $Bi^{5+}$ (12.43e) obtained from $Bi_2O_3$, $Ag_2BiO_3$ and $NaBiO_3$ respectively. If the Bi - O bonds in BBO were to be purely ionic, then the charges of Bi1 and Bi2 should be 13.10e and 12.43e corresponding to $Bi^{3+}$ and $Bi^{5+}$ charge states respectively. The Bi1 and Bi2 charges in the bulk (unstrained) BBO for the I2/m phase are 12.97e and 12.51e respectively. By comparing these charges with the 4+ reference state (12.77e), we notice that the deviations of Bi1 and Bi2 from the reference are greater than 0.2e for monoclinic and rhombohedral cases. The strong deviation from the reference charges on Bi1 (4-δ) and Bi2 (4+δ) for all the space group shows the covalent nature of CD. The absolute charges in the two monoclinic and rhombohedral structures are more or less equal and shift to higher values when the strain changes from -2% to +2% respectively. We notice that, for all the non-cubic structures, the charge on Bi1 changes slowly while the Bi2 charge changes rapidly (>0.1e), as a function of strain. We further note that in the cubic phase, the charges on both the bismuth sites Bi1 and Bi2 are shifted towards the +5 reference charge state due to smaller Bi - O bond length compared to other space groups, due to the enhanced participation of both 6*s* and 6*p* orbitals (coefficient $a_2$ in SI). CD in BBO is further quantified by the charge difference between Bi1 and Bi2 (Fig 1c). The CD for cubic case is significantly smaller than the other three structures due to the absence of OTD. This shows that OTD is not a necessary and sufficient requirement for CD and that CD has its origin in OBD. The tendency of both Bi1 and Bi2 to shift towards 5+ charge state is possible only in the positive-CT limit where the probability of holes on the Bi sites is large. But in a negative CT limit, the probability of holes will be higher in the O 2*p* band as the electrons are localized on to the Bi orbital to form a lone-pair. This would result in the Bi1 and Bi2 charges approaching towards the Bi$^{3+}$ reference charge of 13.10e. We notice a strong tendency of charges approaching the 5+ state in the cubic phase, suggesting that it should have a strong positive-CT insulator character than the other three phases. Further, in the all cases, we notice a tendency towards 3+ state when the strain is strongly tensile and 5+ when the strain is compressive, accordingly strong negative and positive-CT characters are expected respectively.

*Table 2: Calculated Bader charges on sites Bi1 and Bi2 under I2/m, P2$_1$/n, $R\bar{3}$ and $Fm\bar{3}m$ symmetries.*

| Crystal Structure | Space-group | Strain | Charge (e) | |
|---|---|---|---|---|
| | | | Bi1 | Bi2 |
| Monoclinic | I2/m | -2 | 12.94 | 12.45 |
| | | -1 | 12.95 | 12.49 |
| | | 0 | 12.97 | 12.51 |
| | | +1 | 12.97 | 12.54 |
| | | +2 | 13.01 | 12.55 |
| | P2$_1$/n | -2 | 12.95 | 12.46 |
| | | -1 | 12.96 | 12.49 |
| | | 0 | 12.97 | 12.52 |
| | | +1 | 12.98 | 12.54 |
| | | +2 | 13.00 | 12.57 |
| Rhombohedral | $R\bar{3}$ | -2 | 12.95 | 12.48 |
| | | -1 | 12.96 | 12.49 |
| | | 0 | 12.97 | 12.51 |
| | | +1 | 12.99 | 12.54 |
| | | +2 | 13.00 | 12.56 |
| Cubic | $Fm\bar{3}m$ | -2 | 12.80 | 12.42 |
| | | -1 | 12.82 | 12.46 |
| | | 0 | 12.83 | 12.50 |
| | | +1 | 12.85 | 12.53 |
| | | +2 | 13.06 | 12.55 |

*Electronic structure of bulk BBO.-* BD, CD and the corresponding insulating nature can further be understood by investigating the electronic structure of BBO around the Fermi level (E$_F$). This will provide insights into the importance of covalency of the Bi - O bond, the role of O-2*p* band holes in BD and CD, and will eventually shed light on the nature of CD in BBO whether it is ionic or covalent. The electronic structure of all the unstrained space groups along with the corresponding MO states are shown in fig. 2. In the *I2/m* phase, the band between -10 eV and -12.5 eV below the $E_F$ has a doublet peak with equal intensities corresponding to the lone pair of electrons in 6*s* orbitals of Bi1 and Bi2. The splitting of these two features can be attributed to BD present in the system. Where Bi1 has higher contribution from 6*s* orbital, the weight of Bi2 6*s* is lower and vice versa. The DoS of Bi1 6*s* belonging to the lower intense feature at -10 eV and that of Bi2 6*s* at -11 eV are not equal indicating a covalent CD ($4 \pm \delta$). The kink at -12.5 originates due to $\sigma_{1g}$ MO state (fig. S4). We recall from the many-body picture that the lone-pair state with two electrons is higher in energy than the $Bi^{5+}$ state which is hybridized with O - 2*p* orbitals. Therefore, in BBO the ground state is governed by the condition $\boldsymbol{V - U_1 > \Delta}$ and $\boldsymbol{U_2 - U_1 > 2\Delta}$, corresponding to state $|3\rangle$. The almost flat band near -12.5 eV is contributed by Ba 5*p* orbital along with Bi 6*s* and O 2*p*. The bands lying between -6 and -4.5 eV are the $\pi_u$ states corresponding to Bi 6*p* – O 2*p* hybridization with minor contribution of Bi1 and Bi2 6*s*. We further notice that the 6*s* lone-pair band originating at -10 eV disperses towards the $\pi_u$ states, indicating the possibility of Bi 6*sp* - O 2*p* hybridization under strain (Fig. S5). The energy range -4.5 eV to -2.0 eV is dominated by non-bonded oxygen 2*p* states. The band immediately below the $E_F$ is mainly composed of Bi1 6*s* and O 2*p*, with a minor contribution from Bi2 6*p* and 6*s* orbitals, which can be mapped to the antibonding states $\sigma_{1g}^*$ of the MO picture. The bottom of the conduction band predominantly contains Bi2 6*s*, Bi1 6*p* and O 2*p* characters. The states at 5.1 eV above the $E_F$ is mainly Bi1 6*p* with minor contribution from Bi2 6*p* and O 2*p* and corresponds to the Bi 6*p* - O 2*p* hybridized antibonding $\pi_u^*$ MO. Further, there are minor contributions of Bi1 6*s* and Bi2 6*p* orbitals in CB and VB respectively in the vicinity of $E_F$. All these observations indicate that the Bi2 site has more hole than the Bi1 site and CD in BBO is partial.

The electronic structures of BBO under the *P2₁/n, R$\overline{3}$* and *Fm$\overline{3}$m* space-groups are presented in fig. 2. In the monoclinic and rhombohedral phases, the Bi1 6*s* and O 2*p* contributions to DoS below $E_F$ are more or less equal, suggesting that these three phases could be at the boundary of positive and negative-CT regimes. However, in the cubic case the O 2*p* contribution is almost negligible. The top of VB is composed predominantly of Bi1 6*s*, while the bottom of CB has both Bi 6*s* and O 2*p* contributions. This can further be corroborated with the Bader charges of the unstrained cubic which is shifted more towards $Bi^{5+}$ reference charge. This suggests that the lone-pair formation is not favourable in the cubic phase. Interestingly, the intensities of lone pair doublet feature are not equal in the *P2₁/n* phase suggesting a strong asymmetry in the CD ($4 + \delta_1$ and $4 - \delta_2$) due to unequal hole distribution in *a-b* and *a-c* planes (fig. S3). Thus, participation of both Bi 6*s* and 6*p* – O 2*p* in hybridization gives results in Bi 6*sp* – O 2*p* hybridized states in the VB and CB giving rise to *s-p* type charge fluctuations and a positive-CT behaviour, as observed in the bulk cubic phase. On the contrary, if *only* Bi 6*p* hybridizes with O 2*p*, the VB and CB will be dominated by O 2*p* band, giving rise to a negative-CT insulator or *p-p* (O 2*p* – O 2*p*) charge fluctuation.

*Strain dependence of the electronic structure.-* In fig. S5-S8, we show the effect of strain on the electronic structure of BBO for all the space group. The overall bandwidth of Bi 6*s*/6*p* – O 2*p* bands (-13 eV to $E_F$) decreases as we go from -2% to +2% strain. The $\sigma_{1g}$ band intensity also increases with compression, as the overlap of lone pair band and $\pi_u$ states increases. At the same time the intensity of the lone pair band decreases. This overlap is clearly seen at about -6 eV at the Γ point suggesting an enhanced *s-p* charge fluctuation over *p-p* charge fluctuation. Similar behaviour of electronic structure has been seen in one of the previous pressure dependent studies but failed to note the origin of the overlap [25]. The change of the bandwidth of the first Bi 6*s* – O 2*p* feature below $E_F$ also indicates the modification in the charge distribution with strain due to the change in hybridization, effectively modifying δ. As we elongate the lattice, the VB at B–X becomes almost flat, at the same time the CB dispersion increases between A-Γ for *I2/m*. These observations indicate that under compression the dominance of *s-p* charge-fluctuation leads to increase in the probability of $4 + \delta$ charge state or the coefficient $a_2$, while the *p-p* charge-fluctuation increases and stabilizes $4 - \delta$ charge state ($a_1$ dominates over $a_2$) when the lattice is elongated. The strain evolution of the electronic band structure and PDoS of the remaining three phases is qualitatively similar to the *I2/m* phase.

*Role of negative charge-transfer energy (Δ) in CD and BD.-* In presence of strong *e-ph* interaction, BBO was previously suggested to be a Peierl's insulator [47,48]. Alternatively, *e-e* correlation driven negative-CT character in the strong covalent regime has also been discussed in the literature [20,21]. However, in our present study, we have shown that the Bi-6*s* orbital contribution near $E_F$ is higher than O-2*p* orbital, suggesting that BBO does not support strong negative-CT character rather it has an *s-p* type charge fluctuation in the positive-CT limit (as schematically illustrated in Fig. 3). We also divulge that a negative-CT character can be induced and controlled by introducing large tensile strain.

In order to further bolster our argument that the *s-p* fluctuation is dominant in BBO we compare and contrast the electronic structure of BBO with that of $NaBiO_3$ (NBO) reference, in which Bi known to stabilize in 5+ formal charge state and does not show any CD or BD [49]. Figs. 4 (a) and (b) show the calculated PDoS and band-structure of *I2/m* phase of unstrained BBO and $R\bar{3}$ phase of NBO. The states around -8 eV are mainly contributed by Bi-6*s* orbitals of two alternative sites with equal intensity. The position of the Bi 6*s* lone pair orbital in NBO is shifted to higher energy (-8 eV) than in BBO (-10 eV to -12.5 eV) indicating a very weak hybridization between Bi 6*s* and O 2*p* in the former. Interestingly in NBO, the VB is dominated by O - 2*p* contribution, divulging its strong negative-CT character in sharp contrast to BBO which shows a major Bi 6*s* orbital contribution. Also, the DoS of NBO between -2.1 to -4.1 eV has strong Bi 6*p* and O 2*p* mixed character revealing dominant Bi 6*p* - O 2*p* hybridization in this system. These results strongly point towards the conclusion that *I2/m* phase of BBO is a positive-CT insulator with *s-p* type charge fluctuation and NBO is a negative-CT insulator with *p-p* type charge fluctuation. In the completely negative-CT regime as in NBO, the holes will be distributed on the oxygen 2*p* band and not on the bismuth sites, resulting in no CD and BD.

*Valence band spectroscopy.-* To corroborate these theoretical observations, we have also prepared thin films of BBO on two different substrates, STO (001) and NGO (110) so as to contrive different lattice strain, -2% and -1% respectively (fig. S9). The complete description of the method and spectral features (a-f) is presented in the SI file. The valence band spectra (VBS) of the two films (Figure S10) show zero intensity at $E_F$ indicating insulating nature for both the films. The first feature just below $E_F$ at -3eV is attributed to antibonding Bi 6*sp* - O 2*p* hybridized state ($\sigma_{1g}^*$) (Feature a). There is a considerable decrement in the spectral weight in this feature in BBO$_{STO}$ relative to BBO$_{NGO}$ film, consistent with our convoluted total DoS obtained from DFT calculations for similar strain values. The enhancement of Bi 6*sp* - O 2p hybridization is seen as the $\sigma_{1g}$ state at -12.6 eV in BBO$_{STO}$ which is obscure in BBO$_{NGO}$ (Feature f). Further, the lone-pair feature at -10.6 eV is more dominant than $\sigma_{1g}$ in BBO$_{NGO}$ (Feature e) confirming more weight for 4-δ charge state. The charge states of the two films are further ascertained using core level x-ray photoelectron spectrum (fig. S11) which shows dominant 4+δ state for BBO$_{STO}$ and 4-δ state for BBO$_{NGO}$. These experimental results point toward the conclusion that the larger lattice compression enhances *s-p* fluctuation and stabilizes 4+δ charge state while high tensile strain stabilizes 4-δ charge state through *p-p* fluctuation.

In conclusion, we have explored the role of charge fluctuations (*s-p vs. p-p*) on the electronic states of BBO arising due to two competing hybridizations, Bi 6*sp* – O 2*p and* Bi 6*p* – O 2*p*. We have studied the role of strain on the atomic and electronic structure under different crystallographic symmetries. The modification of OTD and OBD changes the Bi - O hybridization which affects the CD and BD. The CD and BD in BBO are coupled. The *s-p* and *p-p* fluctuations favour positive and negative-CT insulating characters respectively. In the monoclinic and rhombohedral phases, the *s-p* and *p-p* fluctuations are comparable, hence they are at the boundary of positive and negative-CT insulating characters. The cubic phase has dominant *s-p* fluctuation or a strongly positive-CT character. Strain can tilt the balance between the two fluctuations; a large tensile strain stabilizes *p-p* fluctuation and drives the system into a negative-CT regime. In bulk NBO, the *p-p* charge fluctuation is dominant and a negative-CT insulating character is observed where the holes are distributed in the O 2p band and hence both CD and BD are absent. The results of this study can provide a good starting point to understand the charge ordering phenomena in other complex oxides like nickelates, plumbates and cobaltates, which have huge relevance in superconductivity, fuel cells and catalysis.

## Acknowledgments

Authors are thankful to DST-SERB for the grants under the projects CRG/2019/001627 and CRG/2021/001021. Dr. D.M. Phase is greatly acknowledged for the fruitful discussions.

## Author contributions

SS performed all the calculations. SS, RR and RJC wrote the draft of the manuscript. All the authors took part in discussion. RR and RJC supervised the work.

# Figures

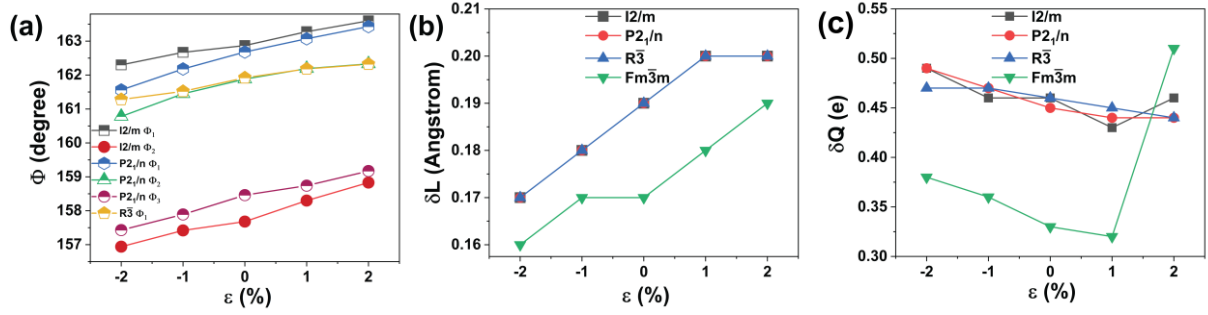

**Figure 1:** *Strain evolution of (a) Bond-angle for I2/m, P2$_1$/n, R$\bar{3}$ and (b) BD and (c) CD of BBO for I2/m, P2$_1$/n, R$\bar{3}$ and Fm$\bar{3}$m structures.*

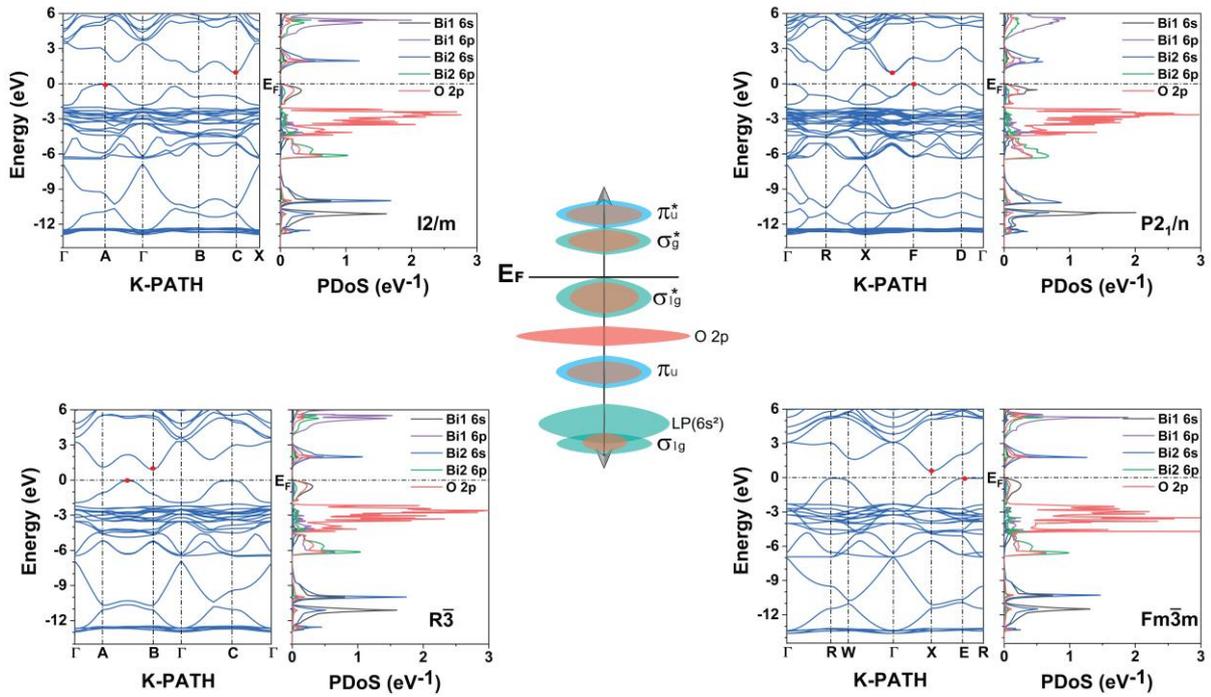

**Figure 2:** *Band structure and PDoS of I2/m, P2$_1$/n, R$\bar{3}$, and Fm$\bar{3}$m phases of BBO and schematic of DoS with the corresponding MO states marked. $E_F$ is marked as dashed line.*

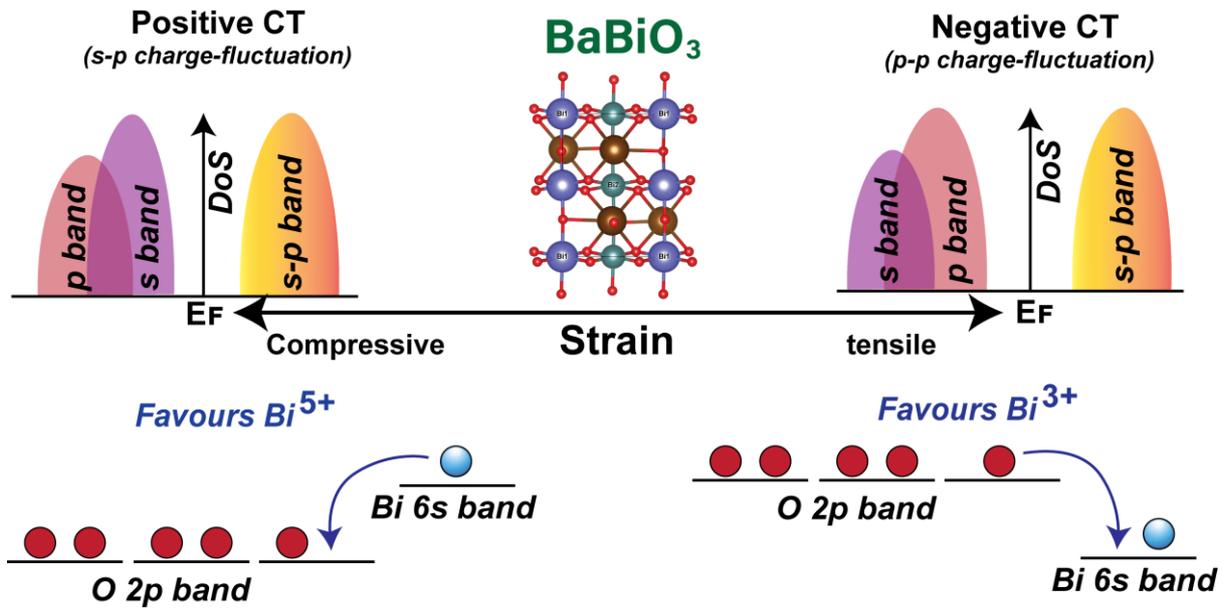

**Figure 3:** *Schematic representation of positive-CT to negative-CT transition under external strain.*

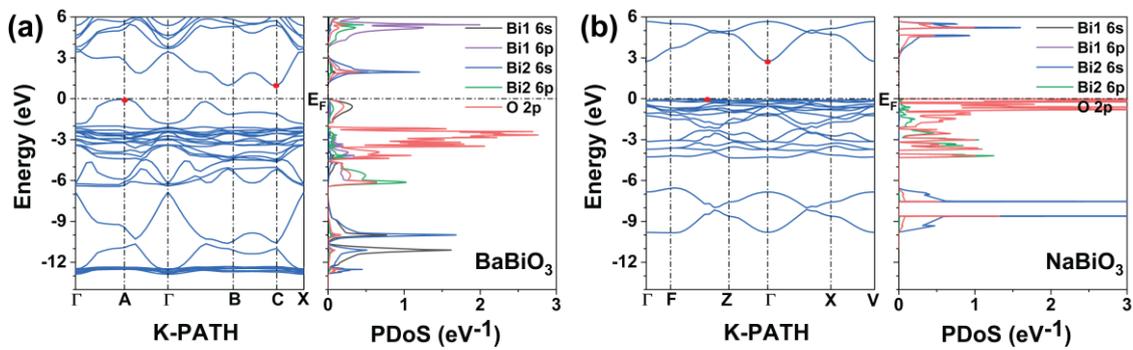

**Figure 4:** *Comparison of band-structure and PDoS of (a) BaBiO$_3$ and (b) NaBiO$_3$.*